\documentclass[12pt,letterpaper]{JHEP3}
\pdfoutput=1
\usepackage{epsfig}

\usepackage{cite}

\def\be{\begin{equation}}
\def\ee{\end{equation}}

\graphicspath{{Figures/}}

\title{Causal Patch Complementarity: \\
The Inside Story for Old Black Holes}

\author{Irfan Ilgin and I-Sheng Yang\\
IOP and GRAPPA, Universiteit van Amsterdam, \\
Science Park 904, 1090 GL Amsterdam, Netherlands
}

\abstract{We carefully analyze the causal patches which belong to observers falling into an old black hole. We show that without a distillation-like process, the AMPS paradox cannot challenge complementarity. That is because the two ingredients for the paradox, the interior region and the early Hawking radiation, cannot be space-like separated and both low-energy within any single causal patch. Either the early quanta have Planckian wavelengths, or the interior region is exponentially smaller than the Schwarzschild size. This means that their appearances in the low-energy theory are strictly time-like separated, which nullifies the problem of double entanglement/purity or quantum cloning. This verifies that the AMPS paradox is either only a paradox in the global description like the original information paradox, or a direct consequence of the assumption that a distillation process is feasible without hidden consequences. We discuss possible relations to cosmological causal patches and the possibility to transfer energy without transferring quantum information.
}

\begin{document}

\section{Introduction and Summary}

The black hole information paradox \cite{Haw76a} has always been an inspiring topic. It is widely recognized that two well motivated believes, purity of Hawking radiation and in-falling vacuum of the horizon, are at odd with each other \cite{Mat09}. Giving up the former leads to information/unitarity loss \cite{Haw76a}, while giving up the later leads to an energy curtain \cite{BraPir09} or a firewall \cite{AMPS}. Other attempts to reconciliate certain aspects of their coexistence always lead to other problems that requires modifications to either quantum mechanics or gravity at low-energy \cite{Gid12,AveCho12,MalSus13,Cho13,Bou13a,LloPre13}. The reason why we have not been actively modifying low-energy effective theories is the concept of complementarity:

%Although the purity of Hawking radiation is difficult to verify in practice, it should not be a reason to ignore this contradiction. A non-standard horizon invalidates general relativity even at the scale of our solar system, and lack of unitarity invalidates quantum physics. Accepting either, we give up the right to use any low-energy physics. We might as well predict that by tomorrow, Earth runs into Venus or LHC explodes. 

{\bf Weak Complementarity:}

{\it The low-energy effective quantum theory, semi-classically coupled to weak gravity, only needs to be self-consistent within individual causal patches.}

{\bf Strong Complementarity:}

{\it The boundaries of causal patches are governed by UV theories (of quantum gravity) which can fix the apparent pathologies in global descriptions.}

\begin{figure}[tb]
\begin{center}
\includegraphics[width=15cm]{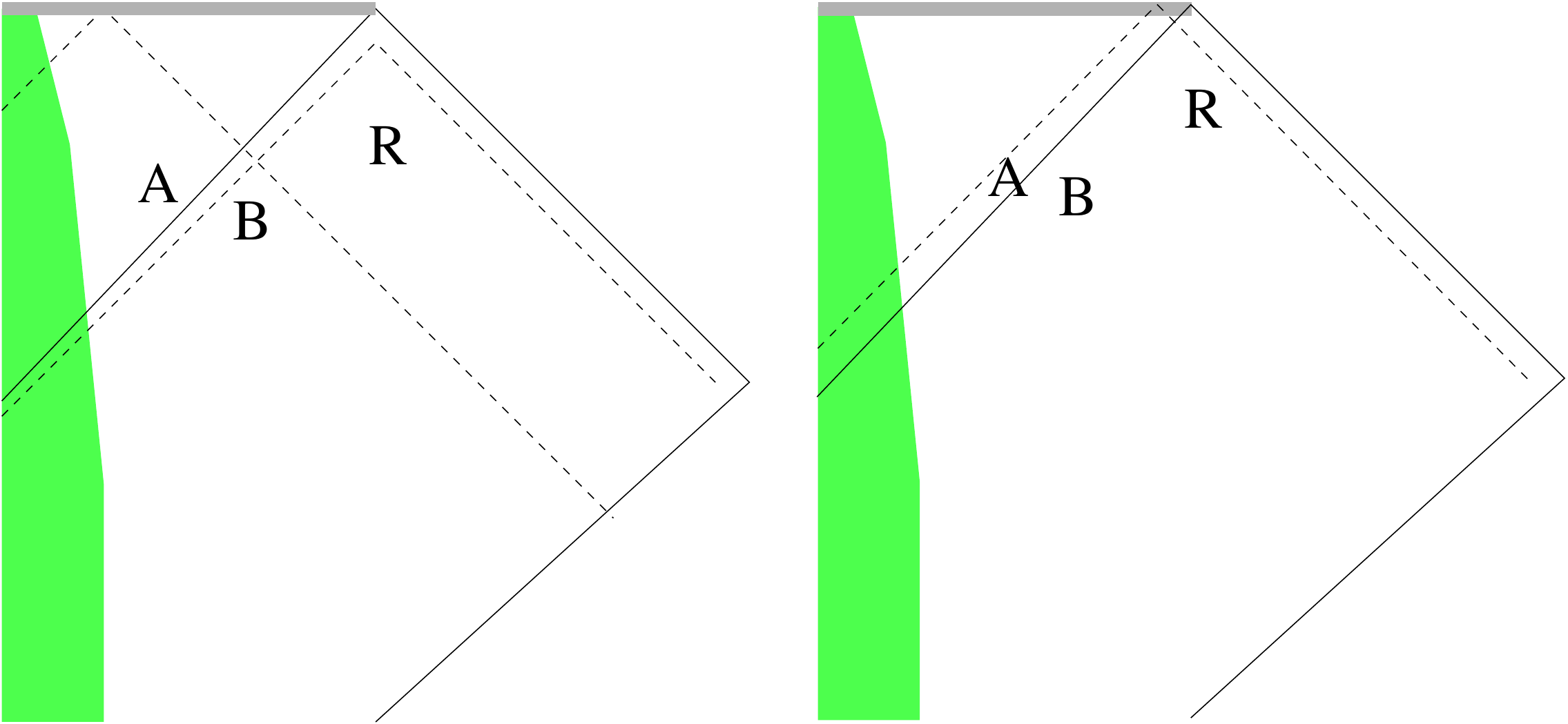}
\caption{The left figure shows the old story for complementarity: an in-falling causal patch needs to confirm that the interior mode $A$ and the exterior mode $B$ are maximally entangled to ensure a normal horizon; an outside causal patch needs to confirm that $B$ is entangled with the rest of Hawking radiation $R$ to preserve unitarity. The right figure shows that in a late, in-falling causal patch, both entanglements need to be confirmed, and that is a paradox for requiring duplicated information in $A$ and $R$.
\label{fig-fit}}
\end{center}
\end{figure}

Stated above is the general form of a global-local complementarity, which might be also useful in cosmology \cite{Sus07,GarVil08,Bou09,BouYan09}. However they are most well studied as the black hole complementarity. As shown in Fig.\ref{fig-fit}-left, the state of the horizon is only described in the casual patch of in-falling observer, while Hawking radiation is in the patch of a distant observer. Since these two potentially contradicting ingredients belong to two different causal patches, weak complementarity guarantees that our everyday application of low-energy effective theory is justified. It remains an interesting topic to find an appropriate statement in strong complementarity, in terms of descriptions of the horizon (and/or the black hole interior) for a distant observer \cite{membrane,Tho90b,SusTho93,Sus93,Ver95,LowPol95,LunMat02,Mat05}. That however belongs to UV physics instead of modifications to low-energy physics.

More than one year earlier, Almheiri, Marolf, Polchinski and Sully (AMPS) formulated a challenge against the status quo \cite{AMPS}. They argued that for a sufficiently old black hole, the information paradox can be revived even abiding the standard of weak complementarity. This new challenge has been paraphrased in many different versions \cite{Bou12,AMPSS,AveCho12,MarPol13,Bou13,Bou13a,Cho13}, and many of them involve tricks like using a distillation process and/or a boundary CFT dual. It is very important to clarify the situation being which one of the following: 
\begin{itemize}
\item Those tricks help to make the paradox clear. \\
In this case there should be a more passive and pristine version of the paradox that demonstrates the core of the problem.
\item Those tricks are essential to establish the paradox. \\
In this case those tricks need more scrutinizations, since the paradox might be an artifact of our na\"ive idealization of those tricks.
\end{itemize}

In this paper, we will ask a well-defined question to clarify the above situation. {\it Without any distillation process, does low-energy theory in individual causal patches run into any pathology?} The na\"ive answer seems to be {\it yes}. It comes from an observation that the two causal patches shown in Fig.\ref{fig-fit}-left are extreme cases: either one jumps into the black hole right away, or one stays outside forever. As shown in Fig.\ref{fig-fit}-right, we can find a generic causal patch between those two. It belongs to an observer who stays outside long enough but eventually falls in. Note that the purity of Hawking radiation has to be verified when the black hole is more than half evaporated, and by that time the black hole (interior) can still be large. Therefore such a late, in-falling causal patch seems to include both ingredients for the paradox. 

\begin{figure}[tb]
\begin{center}
\includegraphics[width=15cm]{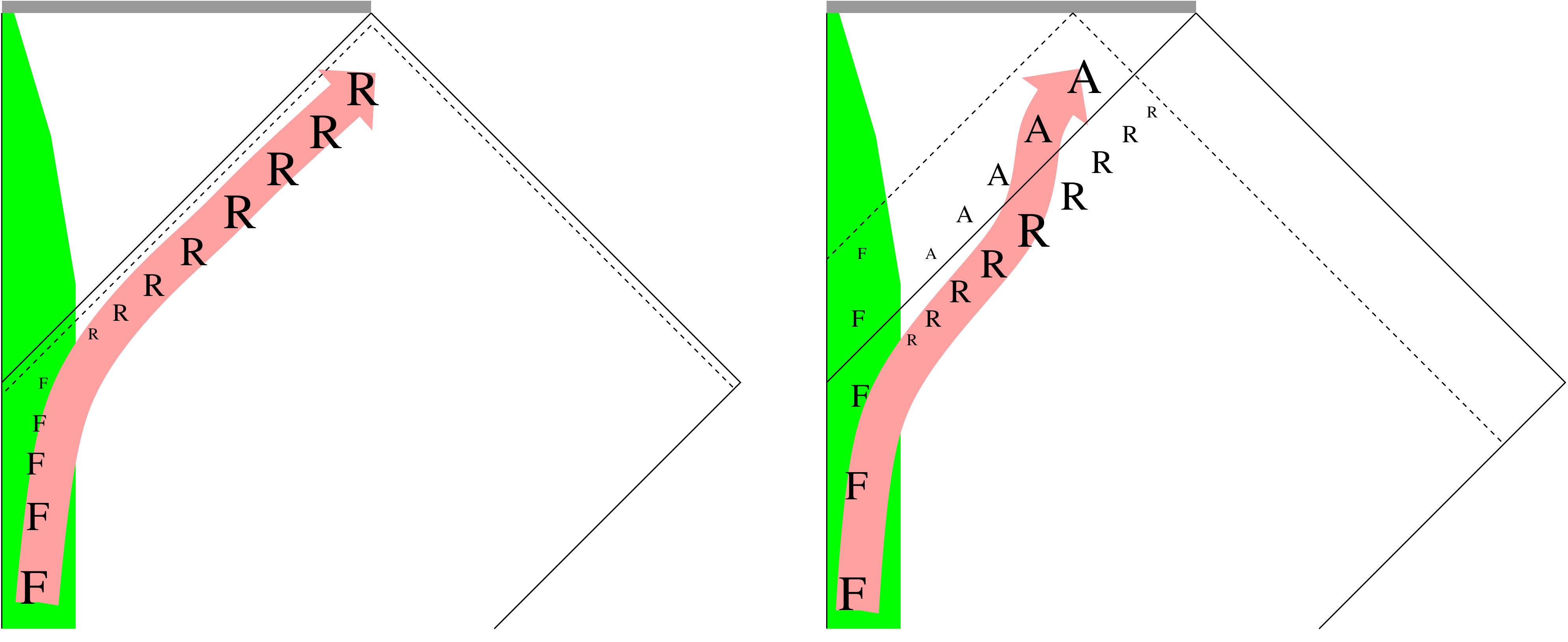}
\caption{The left panel shows the physical interpretation of the outside causal patch, and the right panel shows that for a late, in-falling causal patch (causal patches are bounded by the dotted lines). The size of the letters ($F$, $A$ and $R$) represents the ``wavelength'' of the corresponding physical quantities (in-falling matters, interior modes and Hawking radiation) according to the description within the causal patch. Large letters represent low-energy quantities and small letters represent UV quantities. The pink arrows represent the information flows, which only need to follow low-energy quantities in the usual way. When the carrier of such information becomes UV quantities, unknown UV processes can direct the information to the appropriate places to avoid paradoxes. For the outside patch, such UV flow of information resolves the information-loss paradox. For the late, inside patch, another such UV flow resolves the AMPS paradox.
\label{fig-flow}}
\end{center}
\end{figure}

We will go beyond this na\"ive answer and examine this late, in-falling causal patch in more details. In particular, we emphasize that merely ``fitting into a causal patch'' is not good enough. Otherwise, we do not even need the AMPS argument to have a paradox. The very ancient information paradox uses the in-falling matter and Hawking radiation as its two conflicting ingredients, and they do coexist in the outside causal patch as shown in Fig.\ref{fig-flow}-left. The reason why such situation cannot qualify as a paradox for weak complementarity is because in the outside causal patch, those two ingredients are never {\it space-like separated and both low-energy}.

In Sec.\ref{sec-rules}, we will explicitly state our standard for two physical quantities to ``fit into a causal patch as space-like separated, low-energy quantities''. Note that energy is frame dependent, but whether there exists at least one frame satisfying this standard is a frame independent property of the causal patch. We will show that if such standard is not upheld, then no single observer can properly observe them both. Therefore, a potential paradox built from these two quantities is invalidated by the spirit of complementarity. Note that in \cite{LowTho06}, the integrated boost between two physical quantities was proposed to determine whether they can legitimately form a paradox. That standard can still be defined globally for quantities cannot fit into one causal patch. Our standard is totally within a causal patch and directly related to their simultaneous-observability\footnote{Despite this major difference, if we first limit ourselves to one causal patch and apply the integrated boost standard, then it qualitatively agrees with our standard in the case of a Schwarzchild black hole.}.

In Sec.\ref{sec-Sch} we apply the above standard to a Schwarzschild black hole. We explicitly show that within a late, in-falling causal patch, it is impossible for the interior mode and the early quanta to be space-like separated and both low-energy. Either the early Hawking radiation has Planckian wavelength, or the interior region has a size exponentially smaller than the Schwarzschild radius $\sim M$. We then generalize this result in Sec.\ref{sec-gen} to include possible operations on the early Hawking quanta, including confining them into a box and further thermalization. We find that our conclusion remains unchanged. Within a late, in-falling causal patch, either the black hole interior is exponentially small, or the information within early Hawking radiation goes beyond low-energy physics.

A more dynamical picture of our result is that within low-energy physics, the early Hawking radiation only exists in the early times, and the interior mode only exists in the late times. Since they are strictly time-like separated, the double-entanglement (quantum-cloning) problem does not apply. On every time-slice, this also provides a clear distinction between the ``recent'' Hawking quanta $B$ that live in the low energy theory, and the ``earlier'' Hawking quanta that form $R$ and hide in the UV. If one wishes to think about an $A=R$ map, it can be unambiguously only applied to the UV quantities $R$ but not to the low energy $B$, which avoids the frozen vacuum problem \cite{Bou13,Bou13a}.

In Fig.\ref{fig-flow}, we put the conceptual information flow of this ``inside story'' side-by-side with the well-known ``outside story'' of black hole complementarity and observe their similarity.

{\bf Outside Story:} In the causal patch of an outside observer, the information initially follows the collapsing matter (or anything thrown in later). When the information flows to the boundary, it is transferred to the UV physics. Instead of leaving this causal patch, the UV physics keeps the information in the Planckian neighborhood of the boundary, and later returns it through Hawking radiation. Note that here the boundary of the causal patch is exponentially close to the black hole horizon. Sometimes it is mistaken that the special UV property only occurs for black hole horizon because it is a special place. By the spirit of complementarity, it is the boundary of the causal patch that is a special place for the theory within.

{\bf Inside Story:} In the causal patch a late, in-falling observer, the first half of the story is the same. Note that the causal patch boundary here is also exponentially close to the black hole horizon, although from the inside. Later, Hawking radiation will approach the outside boundary of this causal patch, which means that it again belongs to UV physics. From there on, we are free to claim that the unknown UV physics guides the information in unexpected ways. That is a good news, because strictly after the early Hawking radiation flows to this UV zone, the interior mode $A$ emerges and demands that information. We can simply claim that through unknown UV physics, the information flows there.

In both stories above, we can see there is no need to modify low-energy physics.
\begin{itemize}
\item The information flow is time-like, so there is no need of non-localities \cite{Gid12}.
\item The information flow is future-directed, so there is no need of final-state quantum mechanics \cite{LloPre13}.
\item There is no duplicate information, so there is no need of firewalls \cite{AMPS}.
\item Everything happens within the standard black hole geometry, so there is no need of Einstein-Rosen bridges \cite{MalSus13}.
\end{itemize}

Thus, we have verified that a pristine, distillation-free version of the AMPS paradox does not exist. The pathology explicitly resides in the unknown and idealized distillation process. In Sec.\ref{sec-dis} we point out a possibility to address the remaining distillation problem with our approach. We also discuss possible implications on cosmological horizons.

\section{Weak Complementarity}
\label{sec-rules}

In Sec.\ref{sec-exp} we present a generic way to describe physics within a causal patch and provide the standard for a double-entanglement (quantum-cloning) paradox within the low energy theory: ``{\it Within this causal patch, if there are no space-like surfaces on which both copies of information are carried by low energy quantities, then it is illegal to form a paradox with them.}

In Sec.\ref{sec-fit} we show that our standard is directly connected to whether a single observer can observe any contradiction. We provide pictorial examples to show that our standard has no unphysical side-effects. It does not invalidate paradoxes in usual situations. It only intervenes when the causal structure of the problem obstructs the practical observability of the paradox.

\subsection{Theory within a causal patch} 
\label{sec-exp}
{\bf Definitions}
\begin{itemize}
\item A causal patch $\mathcal{C}$: the entire space-time region within the past light-cone of a point.
\item A foliation $\mathcal{F}(\lambda)$: one parameter family of 3-dimensional space-like surfaces such that every $\mathcal{F}\subset\mathcal{C}$ and $\bigcup_t \mathcal{F} = \mathcal{C}$. In addition, for all $\lambda_i>\lambda_j$, every future directed path from every point in $\mathcal{F}(\lambda_j)$ passes through $\mathcal{F}(\lambda_i)$, and every past directed path from every point $\mathcal{F}(\lambda_i)$ passes through $\mathcal{F}(\lambda_j)$.
\end{itemize}

\begin{figure}[tb]
\begin{center}
\includegraphics[width=12cm]{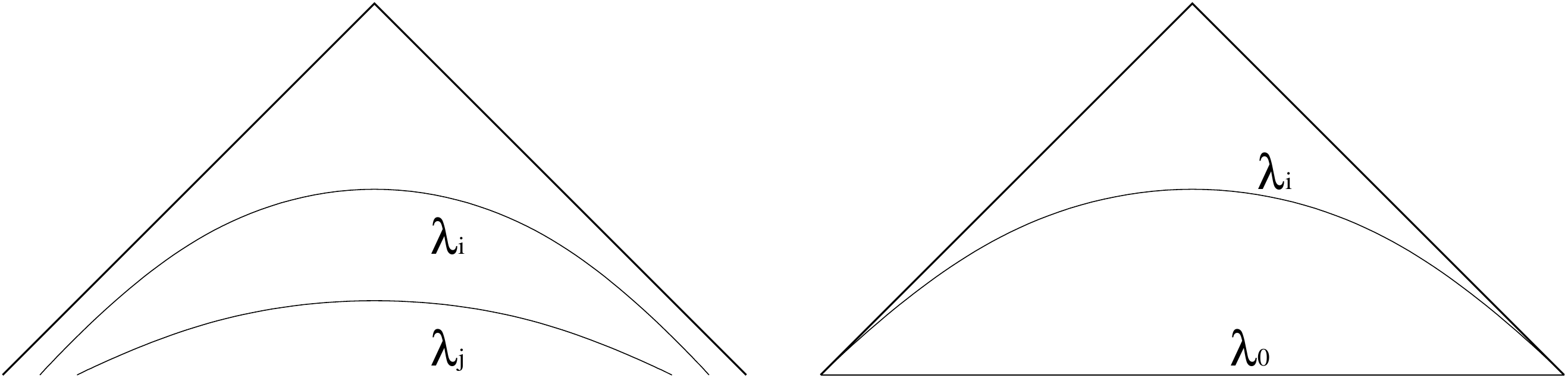}
\caption{Examples of foliations within causal patches. The left panel shows two slices of the standard, semi-infinite hyperbolic foliation given by Eq.~(\protect\ref{eq-open}). The right panel shows the cut-off version where the evolution starts on the initial condition given in the first slice $\mathcal{F}(\lambda_0)$.
\label{fig-patches}}
\end{center}
\end{figure}

The foliation allows us to describe the dynamical evolution as a closed system. For example, the causal patch of the point $(t=0,\vec{x}=0)$ in Minkowski space is naturally foliated by a family of hyperboloids with constant time-like separations from the tip.
\begin{equation}
\mathcal{F}(\lambda) = \{(t,\vec{x})~|~t<0,~\lambda<0,~t^2-|\vec{x}|^2=\lambda^2\}~.
\label{eq-open}
\end{equation}
Of course, this is not a unique choice. A monotonic map $\lambda\rightarrow\lambda'$ leads to a different foliation $\mathcal{F}'$ that is still legal as long as each slice remains space-like. Instead of extending to past infinity, we can also consider a cutoff. Starting from a space-like surface which sets the initial conditions, the foliation only needs to evolve it forward. These example are depicted in Fig.\ref{fig-patches}.

\ \\
{\bf Standard}

A low-energy effective theory within this causal patch has to be consistent in all possible foliations. In particular, since we have conveniently chosen to view it as a closed system, the evolution should be unitary. It looks like a simple job to establish the AMPS paradox. We just need one special example: one causal patch and one foliation in which energy and curvature stay small, but the evolution violates unitarity or has other pathologies.

\begin{figure}[tb]
\begin{center}
\includegraphics[width=8cm]{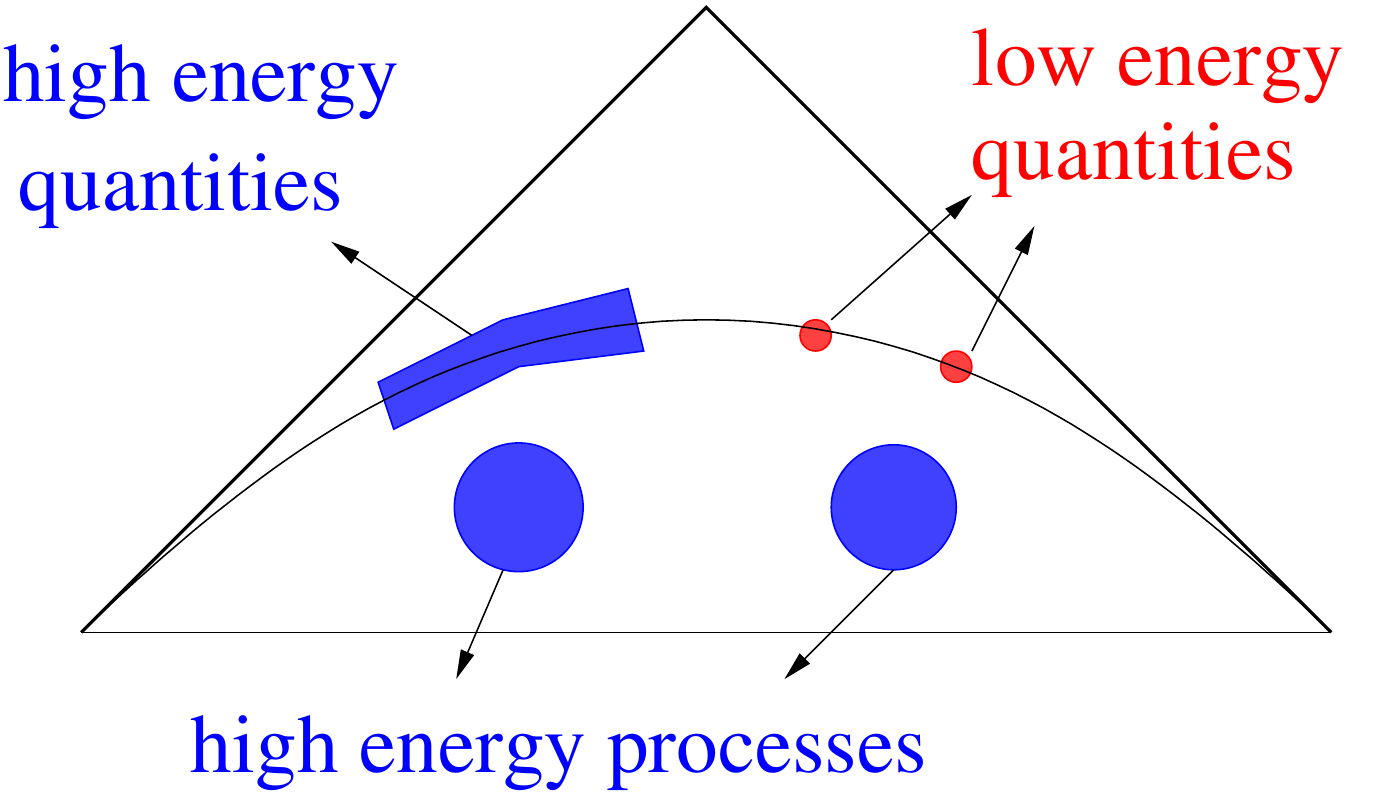}
\caption{Given low-energy input on the initial slice, a consistent low-energy theory can allow high energy events during the evolution or even in parts of the outcome (blue patches). However if the low-energy parts of the outcome (the red dots) show inconsistency, then there is a paradox and the theory needs to be fixed. 
\label{fig-standard}}
\end{center}
\end{figure}

In fact, we will hold an even lower standard for a paradox. We will allow the condition on being ``low-energy'' to be only partially satisfied. First of all there can be gaps in the foliation. As long as one finds two slices $\mathcal{F}(\lambda_i)$ and $\mathcal{F}(\lambda_j)$ that energy is low on both of them, then their states need to be related by a unitary transformation, and the propagation of information should be causal. In between them there might be all sorts of high energy activities like a nuclear bomb or a black hole formation-evaporation, during which even the foliation cannot be consistently defined. We will not use those to disqualify the paradox. Furthermore, we will also allow some regions of the slices to contain high energy quantities. As long as the low-energy regions contain sufficient evidence for pathologies, such as violating the monogamy of entanglement, we accept the existence of a paradox and the necessity to modify low-energy physics. This standard is pictorially summarized in Fig.\ref{fig-standard}.

\subsection{Fitting into the causal patch while being low-energy}
\label{sec-fit}

\begin{figure}[tb]
\begin{center}
\includegraphics[width=12cm]{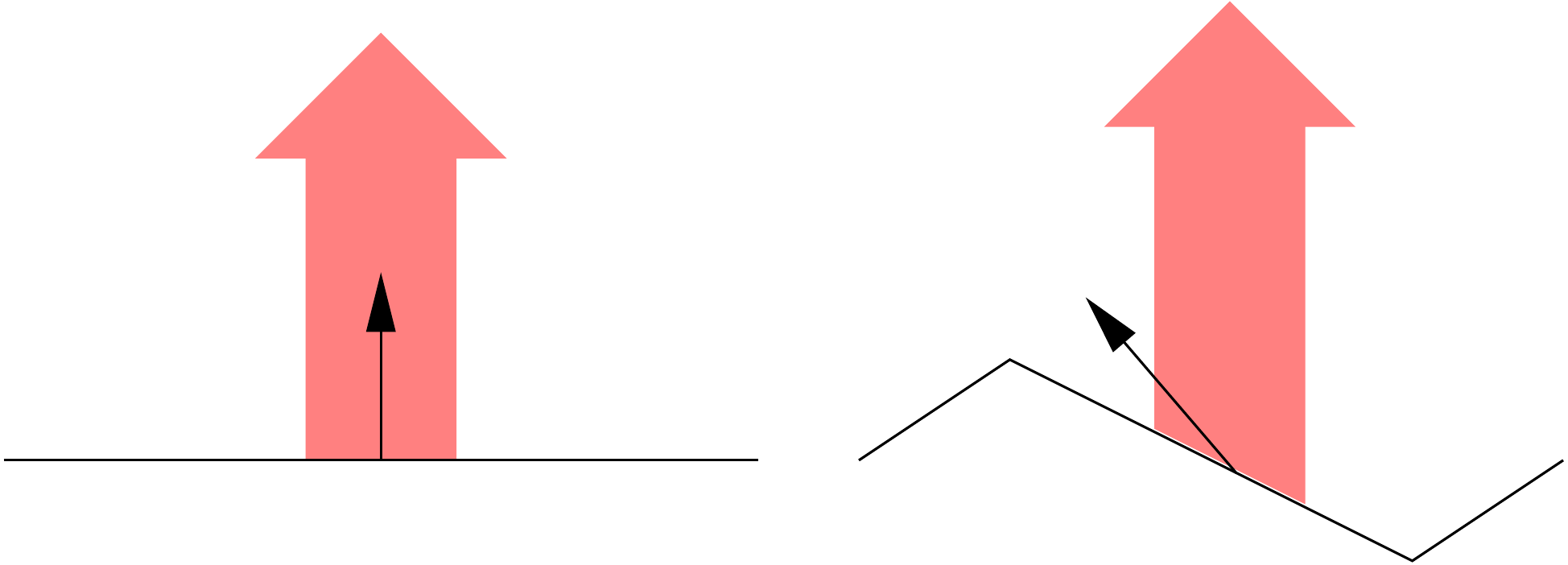}
\caption{In the left panel shows, the big red arrow going up shows represent a low-energy physical quantity and the time-like vector of its natural frame. The arrow is aligned with the time-like vector specified by the foliation (the small black arrow going up), so such physical quantity is low-energy in the foliation. The right panel shows the same quantity in a different foliation, in which there is a large relative boost between the two arrows. Due to such boost, the same quantity appears to be high energy on this foliation.
\label{fig-zigzag}}
\end{center}
\end{figure}

Note that ``energy'', or the length scale, is a frame-dependent quantity. Naturally we define it as the inner product between the momentum vector of a particle and the 4-vector normal to the spacelike surface, so it depends on the choice of foliation. As shown in Fig.\ref{fig-zigzag}, a low-energy quantity in one foliation can become high energy on a different one. It is very reasonable to question the physical meaning of such frame-dependent standard. Indeed the properties of one foliation has no reason to have a deep physical meaning. However, it is a frame-independent physical fact if in all possible choices of foliations, something never happens. 

In other words, we should try our very best to find foliations that makes things as low energy as possible, and see that sometimes it is just impossible. In this section, we will go over pictorial examples to demonstrate when the requirement of fitting a foliation into a causal patch is very restrictive. The remaining freedom can be insufficient to make relevant physical quantities (for the paradox) appear as low-energy. We will then point out that in those situations, these physical quantities in-principle cannot be observed together. This shows that our standard in Sec.\ref{sec-exp} is closely related to observability.

\begin{figure}[tb]
\begin{center}
\includegraphics[width=12cm]{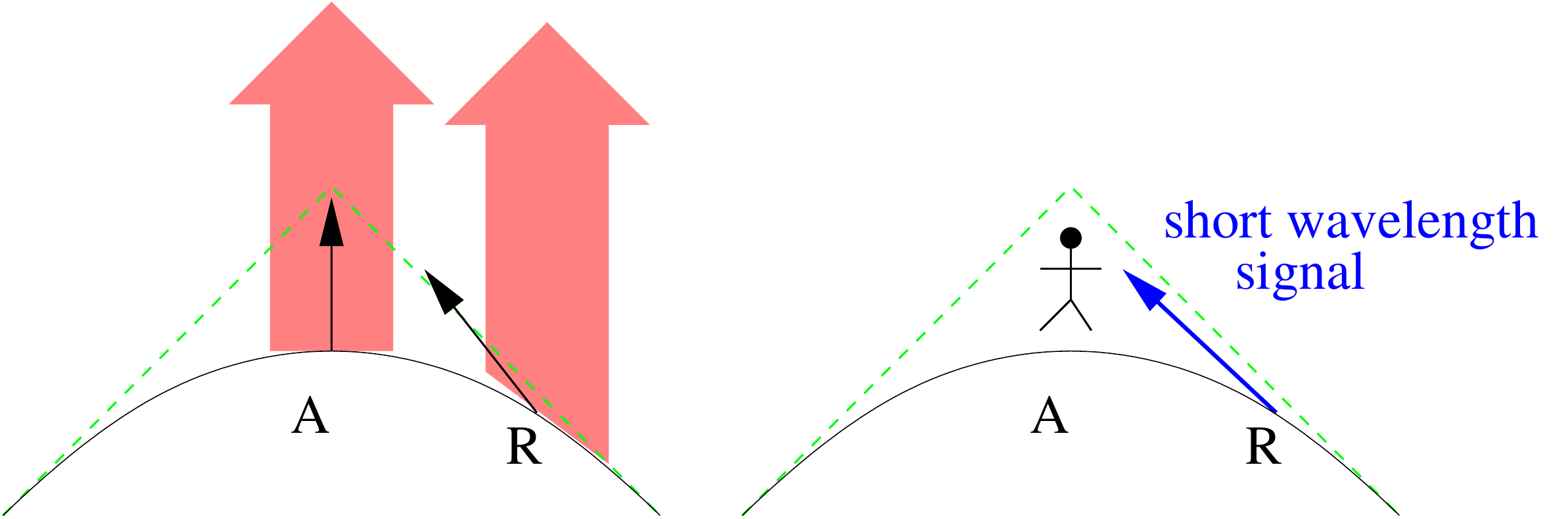}
\caption{$A$ and $R$ are two subsystems on a slice of the foliation within this causal patch (past light-cone in dashed green lines). It may be a paradox if they contain the same quantum information. However if $R$ appears to be high energy due to a large relative boost, then any observer whose experience is limited to this causal patch cannot verify the paradox. As shown in the right panel, for an observer who sees $A$, any information from $R$ has to be at ultra-short wavelengths. That is beyond the applicability of low-energy physics.
\label{fig-AR}}
\end{center}
\end{figure}

Consider the situation in Fig.\ref{fig-AR} in which the initial slice of a foliation is evolved to a later slice where we identify two subsystems $A$ and $R$. If $A$ and $R$ contains the same quantum information, then we may try to establish a paradox on the basis of quantum cloning.

First of all, there are some natural frames in which both $A$ and $R$ are low-energy separately. These are most likely their rest frames or the rest frames of the sources if they are radiations. Since $A$ is in the center of the patch and its velocity aligns with the normal 4-vector of the foliation, it remains to be low-energy in this foliation. On the other hand, $R$ is on the edge of the patch and has a large relative boost to the foliation, so it might be high energy. For an observer whose experience is confined within this patch, this situation implies a true limitation. This observer has to be at the top of the patch to see $A$, from where he can only catch a fleeting glimpse into $R$ before it exits the patch. If this glimpse does not last long enough, this observer cannot decipher information from $R$, at least not when he is still inside this patch.

\begin{figure}[tb]
\begin{center}
\includegraphics[width=14cm]{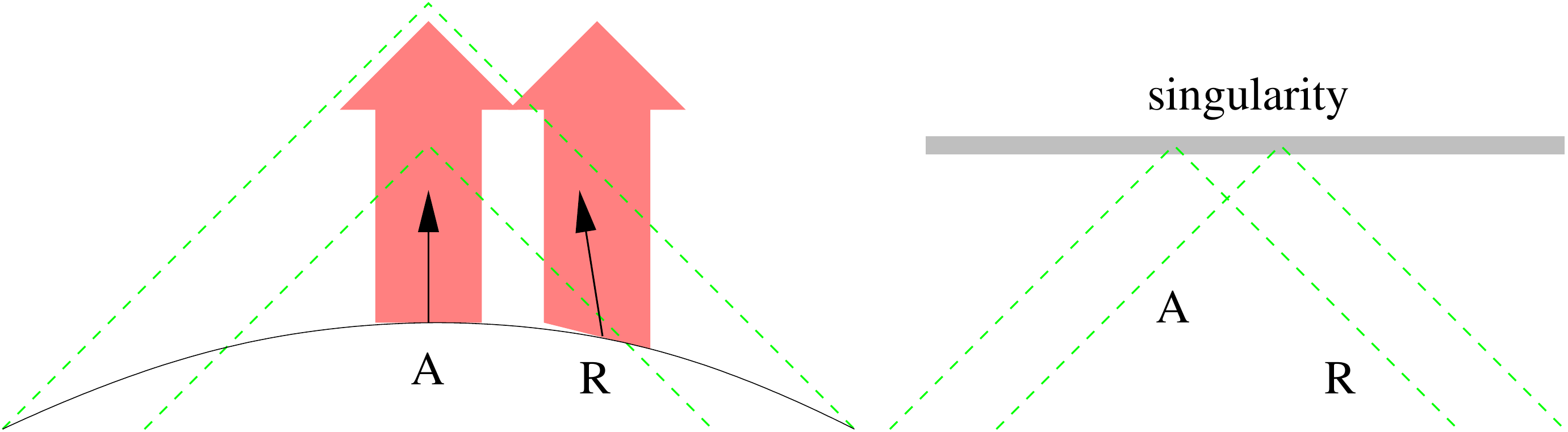}
\caption{The left panel shows that if the causal patch in Fig.\protect\ref{fig-AR} is from an interior point, then one can take a bigger causal patch which totally covers the previous one, and it can easily accommodate both $A$ and $R$ in low-energy. The right panel shows that if the causal patch is from a boundary point, for example at future singularity, then there is no bigger causal patch. Shifting it to the left or right will not be enough to make $A$ and $R$ both low-energy, or even have trouble to include them both.
\label{fig-other}}
\end{center}
\end{figure}

If this causal patch is from an interior point, then we cannot reject the paradox this way. Since the limitation only applies to a confined observer, one can choose a later observer who is less confined. Equivalently, as shown in Fig.\ref{fig-other}, one can take a later (larger) causal patch, and in this new patch $A$ and $R$ are easily both low-energy in some foliation. This is a consistency check that our standard does not reject paradoxes while it should not.

On the other hand, if the tip of the patch is at a singularity, then the limitation is truly meaningful. Now there is no bigger causal patch. One can consider another patch to the right of this one. It can put $R$ at the center but then $A$ will appear to be high energy. In this case no single observer can collect information from both $A$ and $R$, therefore the presumed paradox is not observable and should be rejected. This is also shown in Fig.\ref{fig-other}.

\begin{figure}[tb]
\begin{center}
\includegraphics[width=12cm]{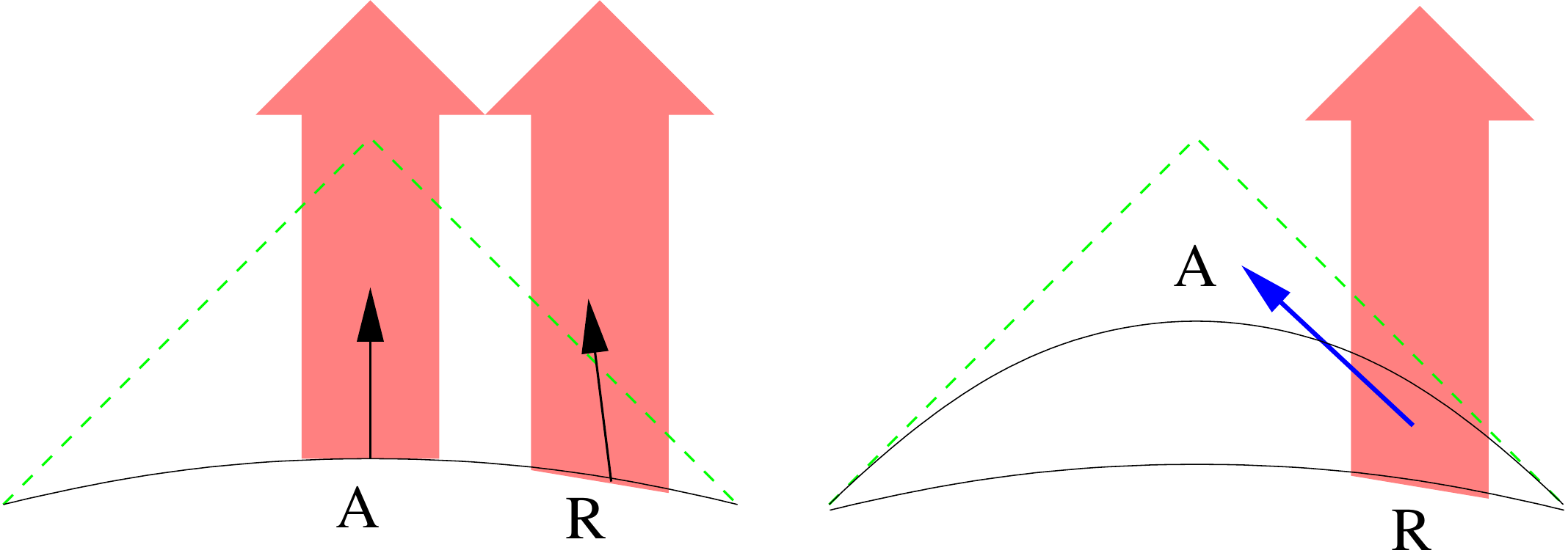}
\caption{The left panel shows the case that $A$ and $R$ appear to be low-energy on an earlier slice. The right panel shows the case that $A$ does not have a low-energy past, while $R$ only becomes low-energy on an earlier slice when it is in the causal past of $A$. This allows the interpretation that $R$ gets burned out at the ``horizon'' and propagates to $A$ (following the blue arrow) through an unknown high energy process. 
\label{fig-burn}}
\end{center}
\end{figure}

In Fig.\ref{fig-burn} we see another consideration that we can evolve both $A$ and $R$ backward in time, and they may appear less boosted with respect to an earlier slice of the foliation. Indeed if $A$ and $R$ together appear on an earlier slice and are both low-energy, that slice suffices to establish the paradox. Later we will see that in the AMPS case, $R$ is low-energy at an early time while $A$ is not, and $A$ is low-energy at a later time while $R$ is not. They are never both low-energy in at the same time. This is a clear sign that we should probably reject the paradox, because the cloning problem only sustains when the full Hilbert space contains the product of their individual Hilbert spaces, $\mathcal{H}_A\times\mathcal{H}_R\subset\mathcal{H}$. This picture is natural only when $A$ and $R$ are space-like separated. If $R$ is in the past of $A$, then it could {\it be} the past of $A$, and it is not a paradox for them to have the same quantum information.

A natural interpretation for time-like separated $A$ and $R$ demanding the same information is simple: $R$ gets burned out at the ``horizon'' (the outside boundary of the patch, not the black hole horizon) and its information content travels to $A$. Note that this is a process in UV physics, so our lack of expectation of such an information flow, or any low-energy setup to block it, cannot be a sufficient reason to insist on the paradox\footnote{We thank Daniel Harlow for pointing this out.}.

\section{Causal patch for an old Schwarzschild black hole}
\label{sec-Sch}

Here we will apply the standard in Sec.\ref{sec-rules} to examine causal patches in the geometry of an old Schwarzschild black hole. We will focus on the two ingredients which are used to form the information paradox in \cite{AMPS}. The first ingredient is the near horizon region of size $\sim M$. This region generates Hawking radiation by separating the pair of interior $A$ and exterior $B$ modes which formed the local vacuum state. When the exterior mode $B$ has a wavelength $\sim M$, then it is officially a out-going Hawking quantum. Due to how it was generated, its state is maximally entangled with its interior partner $A$ also with a wavelength $\sim M$ \cite{SchUnr10}.

The second ingredient is the early Hawking radiation $R$. Because the evaporation process is unitary, an out-going Hawking quantum $B$ should be maximally entangled with $R$. This double entanglement requires $A$ and $R$ to carry duplicated quantum information. So if they both appear within this causal patch and are low-energy quantities on certain slice of some foliation, then it is indeed a paradox for weak complementarity. 

Here we present an explicitly calculation to show that for all possible foliations in all such causal patches, there is no single slice on which $A$ and $R$ coexist as low-energy quantities. 

\begin{figure}[tb]
\begin{center}
\includegraphics[width=14cm]{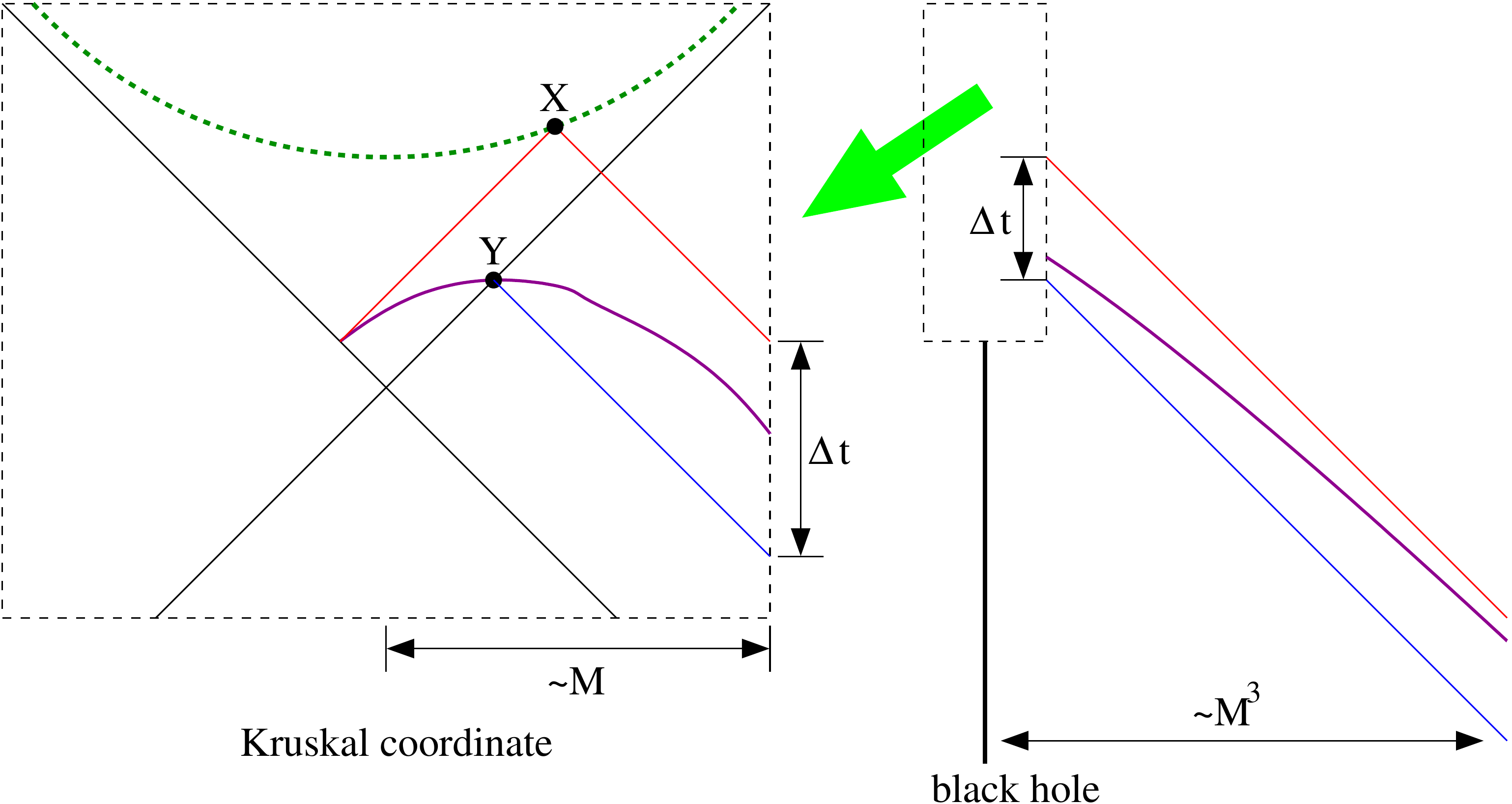}
\caption{The right side shows the outside calculation where the typical length scale involved is $M^3$. The near horizon $\sim M$ region of the black hole is effectively a point. We zoom in to that region (the dashed box) in the Kruskal coordinate. The two anchor points: $X$ is the tip of a causal patch (enclosed by the red past light-cone) on the singularity (the green curve); $Y$ is where the space-like slice (yellow curve) intersects with the black hole horizon. The past light-cone (blue line) from $Y$ is convenient for defining the distance $\Delta t$ between these two anchor points, and the space-like slice must be bounded between these two light-cones.
\label{fig-strategy}}
\end{center}
\end{figure}

Our calculation has two parallel sessions addressing the inside and the outside. Our inside calculation is in the Kruskal coordinates and includes the black hole interior and some outside region near the horizon. Our outside calculation deals with regions far away from the black hole where the early Hawking radiation has propagated to, and we can simply use the Schwarzschild coordinates. This inside-outside separation is illustrated in Fig.\ref{fig-strategy}. There are two important anchors in our calculations. The first anchor $X$ is the tip of the causal patch at the singularity, and the second anchor $Y$ is where the slice intersects the horizon. It is useful to draw the past light-cone from both anchor points as we did. Since the slice we are looking for must be bounded between these two light-cones.

These two light-cones extend through both parts of our calculation. The most convenient way to connect the calculations is through $\Delta t$, the Schwarzschild time difference when these two light-cones intersects the boundary between the two parts of our calculation. In other words, it represents the distance between the two anchors\footnote{Note that there is a hidden spherical symmetry in the diagram, but the anchor points $X$ and $Y$ are only one particular point even in the angular coordinate. In the inside calculation, we are calculating exactly the distance along this particular angular direction. In the outside it looks like we are also only calculating the Hawking quanta flowing along this direction. However, the extension of these light-cones to other angular directions will keep the same separation $\Delta t$, so our outside calculation is actually valid in all directions.}. This turns out to be the only relevant parameter. We will present the detail calculation in the next two sub-sections. Here we briefly summarize the result.

The interior mode $A$ only exist on the segment of the slice to the left of point $Y$. We define $\lambda_A$ to be the proper length of this segment and find it bounded from above by
\begin{equation}
\lambda_A \lesssim M e^{-\Delta t/4M}~.
\end{equation}
Namely, inside this causal patch, we need $\Delta t < M$, otherwise the interior mode is exponentially small.

On the other hand, the wavelength of the early Hawking quanta on this slice (the inverse of its 4-momentun projected onto the normal timelike vector) is bounded by
\begin{eqnarray}
\lambda_R < \sqrt{\frac{\Delta t}{M}}~.
\end{eqnarray}
This means that we need $\Delta t > M$, otherwise these quanta have Planckian wavelengths. Since $\Delta t$ cannot be both bigger and smaller than $M$, either we cannot address the information content in $R$, or we do not describe the near horizon process that generates late Hawking quanta. The two ingredients for the AMPS paradox have failed to coexist in the low-energy theory within a causal patch.

In the situation where we choose $\Delta t\sim M$, the near horizon region of this slice agrees with an in-falling observer free-falling from $\sim M$, so the interior mode $A$ looks normal. However the early Hawking quanta suffer a large boost $\gamma\sim M$ which contracted their wavelengths from $\sim M$ to order one. This shows that our standard qualitatively agrees with the standard of integrated boosted \cite{LowTho06}.

\subsection{Inside: the interior mode $A$}
\label{sec-inside}

\begin{figure}[tb]
\begin{center}
\includegraphics[width=14cm]{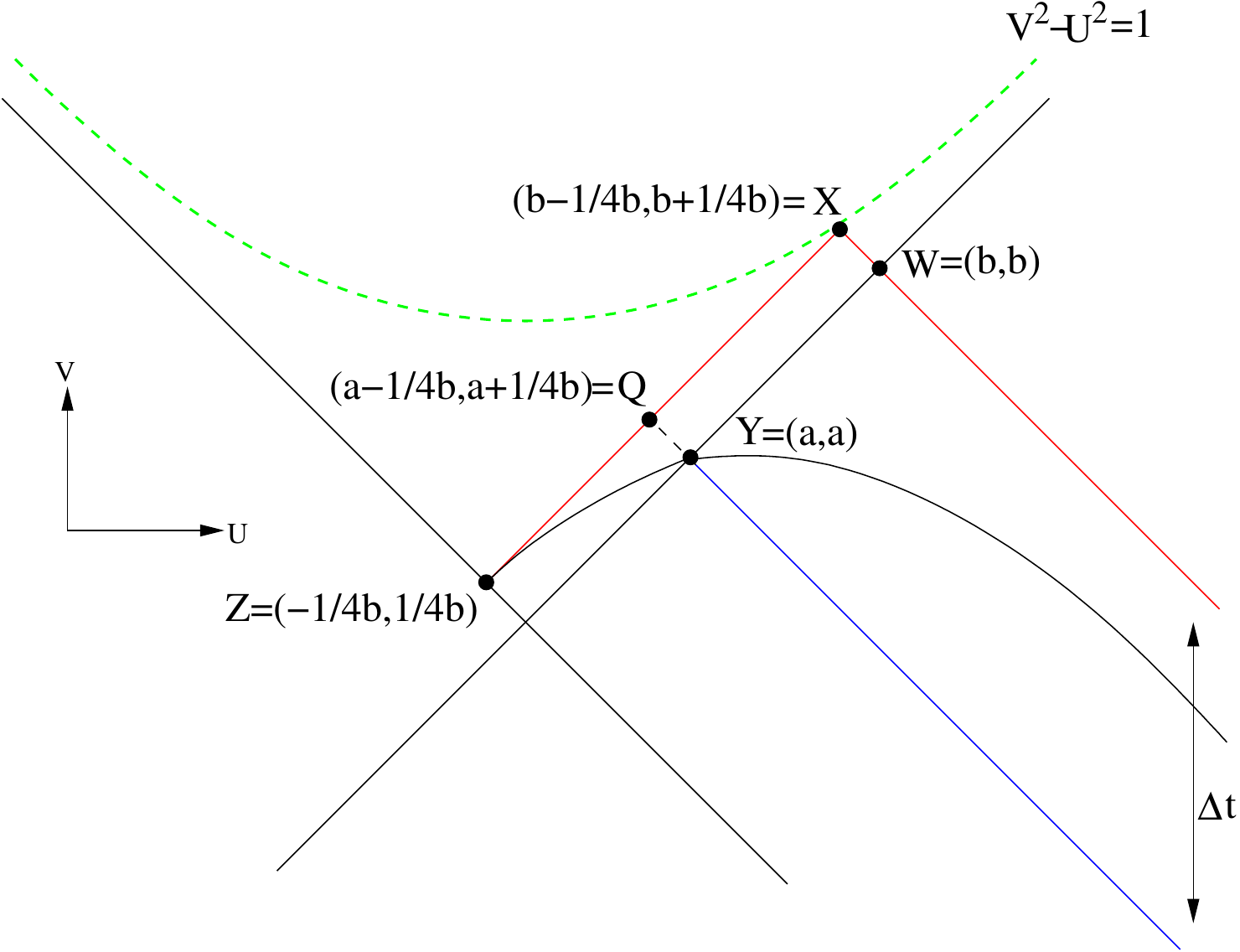}
\caption{The calculation of the length of the interior mode $A$ in the Kruskal coordinates. Points $X$ and $Y$ are the same as those in Fig.\protect\ref{fig-strategy}. The $(U,V)$ coordinates of other important points in the calculation are also shown here.
\label{fig-inside}}
\end{center}
\end{figure}

Consider a half-evaporated black hole with current mass $M$. The metric is usually given in the Schwarzschild form. 
\be
ds^2=-\Big(1-\frac{2M}{r}\Big)dt^2+\Big(1-\frac{2M}{r}\Big)^{-1}dr^2+r^2 d\Omega^2~.
\ee
However to analyze the geometry across the horizon, it is more convenient to remove the coordinate singularity by going to the Kruskal-Szekeres coordinates.
\be
ds^2=\Big(\frac{32M^3}{r} \Big)e^{-r/2M}(-dV^2+dU^2)+r^2d\Omega^2
\ee
This is related to the Schwarzschild coordinates by 
\begin{eqnarray}
V&=&\left(1-\frac{r}{2M}\right)^{1/2} e^{r/4M} \cosh\left(\frac{t}{4M}\right)~,\\ \nonumber
U&=&\left(1-\frac{r}{2M}\right)^{1/2} e^{r/4M} \sinh\left(\frac{t}{4M}\right)~,
\end{eqnarray}
in the interior $(V^2>U^2, r<2M)$, with a singularity $V^2-U^2=1$. In the exterior $(V^2<U^2, r>2M)$, they are related by
\begin{eqnarray}
V&=&\left(\frac{r}{2M}-1\right)^{1/2} e^{r/4M} \sinh\left(\frac{t}{4M}\right)~,\\ \nonumber
U&=&\left(\frac{r}{2M}-1\right)^{1/2} e^{r/4M} \cosh\left(\frac{t}{4M}\right)~.
\end{eqnarray}
The coordinates of the two anchor points are given by
\begin{eqnarray}
X &=& (U_X,V_X) = (b-\frac{1}{4b},b+\frac{1}{4b})~, \\
Y &=& (U_Y,V_Y) = (a,a)~.
\end{eqnarray}
The variable $b$ is chosen such that the past light-cone from $X$ intersects the horizon at $W=(b,b)$. These points and a few that we will define later are shown in Fig.\ref{fig-inside}. The choice of variables $a$ and $b$ simplifies the relation to the important parameter $\Delta t$, which is the difference in the Schwarzschild time if we follow the past light-cones from these two points to some $r>2M$ outside.
\begin{equation}
e^{\Delta t/4M} = \frac{b}{a}~.
\end{equation}

We would like to calculate the length of the interior mode $A$ on a space-like surface which passes through point $Y$ and bounded by the past light-cone from $X$. This length will be bounded by the space-like geodesic distance between point $Y$ and point $Z=(-\frac{1}{4b},\frac{1}{4b})$\footnote{Point $Z$ is on the other horizon of an eternal black hole, which technically speaking does not exist in the geometry of a black hole formed by collapse. However that means before this slice extends to point $Z$, it would have run into the collapsing matter and stopped. Thus the distance between $Y$ and $Z$ is a good upper-bound.}.
\begin{eqnarray}
\lambda_A &<& \int_Z^Y \sqrt{\frac{32M^3}{r}}e^{-r/4M}\sqrt{dU^2-dV^2}
\\ \nonumber
&\leq& \sqrt{\frac{32M^3}{r_Q}}e^{-r_Q/4M} \int_Z^Y \sqrt{dU^2-dV^2} 
\leq \frac{a}{b} \sqrt{\frac{32M^3}{r_Q}}e^{-r_Q/4M}~.
\label{eq-lA1}
\end{eqnarray}
The point $Q=(a-\frac{1}{4b},a+\frac{1}{4b})$ is where the metric factor reaches the maximal value for any space-like path between $Y$ and $Z$.
\begin{equation}
\left(1-\frac{r_Q}{2M}\right)e^{r_Q/2M} = \frac{a}{b} 
=e^{-\Delta t/4M}~.
\label{eq-Q}
\end{equation}
After taking out the metric factor, the remaining integral is simply maximized by a straight line.

According to Eq.~(\ref{eq-Q}), $r_Q$ decreases as $\Delta t$ increases. That means the upper bound in Eq.~(\ref{eq-lA1}) decreases as $\Delta t$ increases. For example, if
\begin{equation}
\frac{\Delta t}{4M} = \ln\frac{b}{a} > \ln 2~,
\end{equation}
then we have
\begin{equation}
\lambda_A < 4\sqrt{2}Me^{-1/4-\Delta t/4M} < 2\sqrt{2}Me^{-1/4}~.
\label{eq-lA}
\end{equation}
So we would like $\Delta t>M$ for this causal patch to contain any interior mode $A$ with $\lambda_A \gtrsim M$.

\subsection{Outside: the early Hawking quanta $R$}
\label{sec-outside}

In the outside we only need to consider the early Hawking quanta, which are $\sim M^3$ away from the black hole. That means we can even ignore the Schwarzschild metric and treat it as flat space. The relative error we are making is $\sim M^{-2}$, which is negligible in the limit of a large black hole.

\begin{figure}[th]
\begin{center}
\includegraphics[width=6cm]{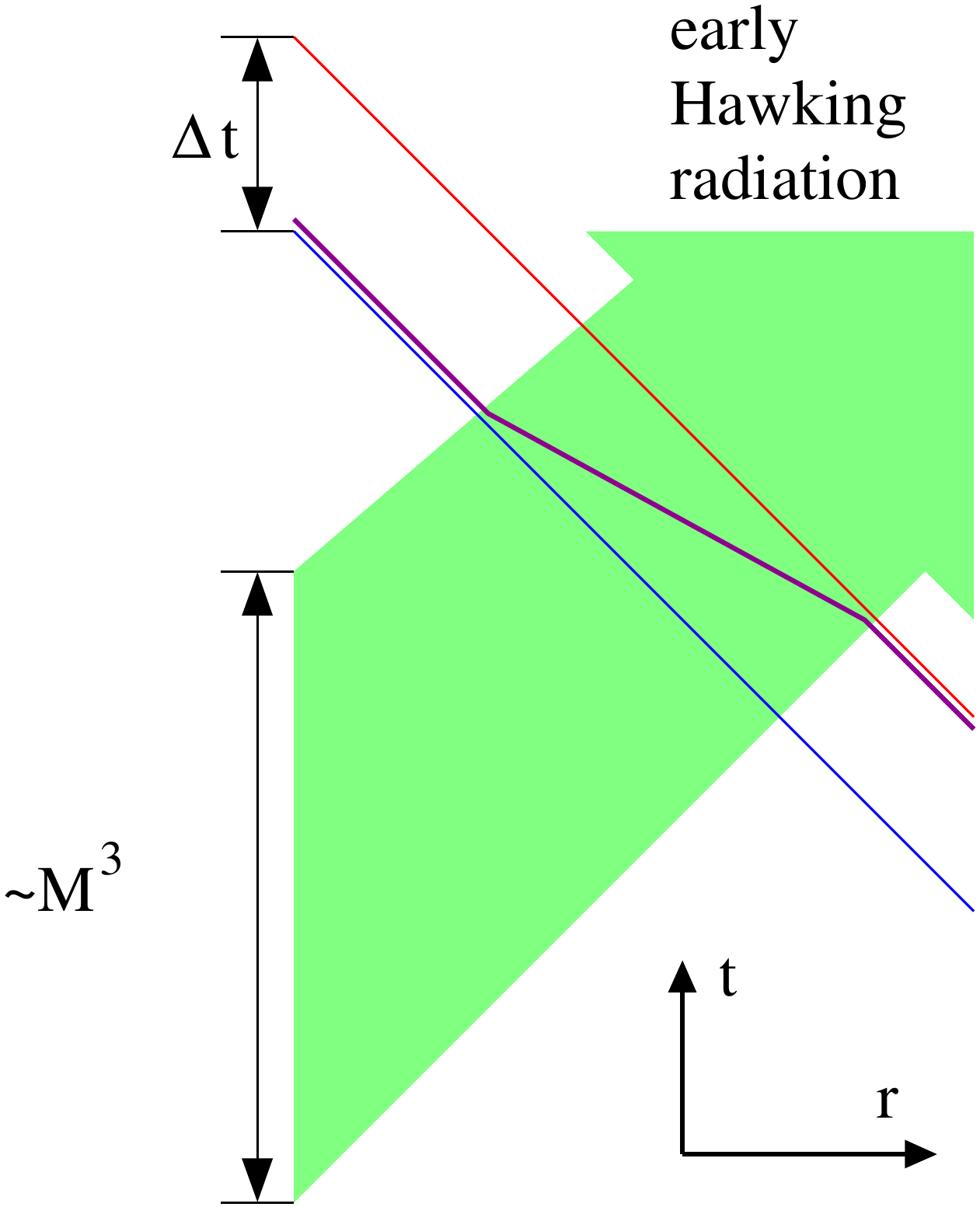}
\caption{The outside calculation. Early Hawking radiation comes from the black hole during a time duration $\sim M^3$ and flows as the big green arrow. The slice (purple) intersects with this flow and is bounded by the blue and red light-cones with separation $\Delta t$. In order to minimize the relative boost between the slice and the Schwarzschild frame (which is equivalent to maximizing the proper length of this segment), it must go through the corners.
\label{fig-outside}}
\end{center}
\end{figure}

Our goal is as shown in Fig.\ref{fig-outside}. The early Hawking radiation will pass through some region of this slice, and the slice is bounded between the two past light-cones. We would like to minimize the energy of the early Hawking quanta in the frame determined by this slice. That means minimizing the relative boost between the Schwarzschild time and the time-like direction orthogonal to this slice. Since we do not care about other regions on this slice which do not contain any early Hawking quanta, the maximum is reached by the slice shown in Fig.\ref{fig-outside}. This boost is given by
\begin{eqnarray}
v &=& \frac{M^3}{M^3+\Delta t}~, \nonumber \\
\gamma &=& 
\frac{1}{\sqrt{1-v^2}}\approx\sqrt{\frac{M^3}{\Delta t}}~.
\label{eq-boost}
\end{eqnarray}
That means the early Hawking quanta on this slice will have wavelength
\begin{equation}
\lambda_R < \frac{M}{\gamma} = \sqrt{\frac{\Delta t}{M}}~.
\label{eq-lR}
\end{equation}
This means that we need $\Delta t > M$ to make $\lambda_R>1$, so the early Hawking quanta are not Planckian. That is in direct conflict with Eq.~(\ref{eq-lA}) which needs $\Delta t < M$ to make $\lambda_A \gtrsim M$.

\section{Generalizations}
\label{sec-gen}

In Sec.\ref{sec-Sch} we showed that the problem $R$ might run into in a late, in-falling causal patch is that the wavelength of individual Hawking quantum becomes the Planck scale. That is a combination of their original wavelength $M$ and the large distance $M^3$ from the black hole. These two things can be modified. For example, Hawking radiation can interact with a cloud of dust and further thermalize into more quanta with longer wavelengths. One can also try to confine the quanta into a box so they are not so far away from the black hole\footnote{Since it is hard to construct perfect reflecting surface (especially for gravitons), the ideal setup is embedding the black hole in an AdS space \cite{HawPag82}.}. In this section we will argue that our conclusion still holds under those changes.

The first issue is that those changes, thermalization and confinement, also change the number (density) of quanta. ``Wavelength should be smaller than the Planck scale'' is the standard for one quantum, and we need to find a generalization of such standard for more quanta. A natural guess is 
\begin{equation}
N \ll \lambda L~.
\label{eq-lowenergy}
\end{equation}
If there are $N$ quanta of wavelength $\lambda$ in a region of size $L$, then it is actually a black hole. In this sense one quantum is the special case that $L=\lambda$.

Eq.~(\ref{eq-lowenergy}) is exactly the standard we will follow. In particular, note that $\lambda$ and $L$ are frame-dependent quantities. We will argue that whether information is accessible is also frame-dependent. This is similar to the examples in Sec.\ref{sec-rules}. If Eq.~(\ref{eq-lowenergy}) cannot be satisfied on all possible foliations with a causal patch, then it means no observer within this causal patch can read such information.

\begin{figure}[th]
\begin{center}
\includegraphics[width=12cm]{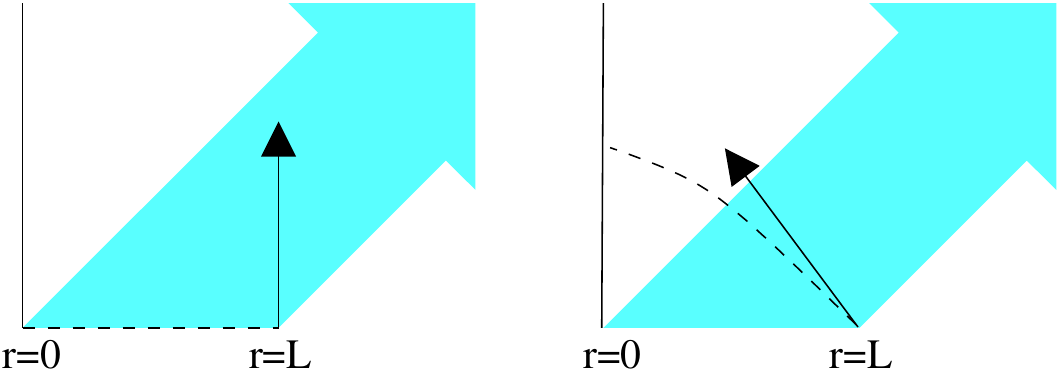}
\caption{Both figures have a suppressed spherical symmetry around $r=0$. Whether the information (blue flow) on a space-like surface (the dashed line/curve) qualifies as low-energy is related to whether observers in the rest frame of such surface (arrows) can read such information when they flow out.
\label{fig-gen}}
\end{center}
\end{figure}

First consider the situation in Fig.\ref{fig-gen}-left: on a usual space-like surface, within a shell of radius $L$, we have $N$ quanta of wavelength $\lambda$. We claim that the information content in this group of quanta is illegal if Eq.~(\ref{eq-lowenergy}) is not satisfied. The obvious reason is that this region actually forms a black hole.

Now take a more observer oriented point of view. Let a group of observers sitting at the shell and read a flow of information from those $N$ quanta. If they find that Eq.~(\ref{eq-lowenergy}) is violated, they should conclude that such information is a lie. Because if there was such information inside the shell, the observers should have collapsed into a black hole with it.

Next we consider the same situation with another space-like surface as in Fig.\ref{fig-gen}-right. As we already saw in Sec.\ref{sec-Sch}, the choice of this kind of surface is enforced by the need to fit into some causal patch and to include another ingredient of the paradox. For simplicity we say that the relevant region (where the information flow through) has a boost $\gamma$ relative to the surface in Fig.\ref{fig-gen}-left.

Our claim is that we should use $\lambda$ and $L$ in the frame of this slice in Eq.~(\ref{eq-lowenergy}) to determine whether the information within these quanta is legal or not. In this case those two length scales are blue-shifted.
\begin{equation}
N \ll \frac{\lambda}{\gamma}\frac{L}{\gamma}~.
\label{eq-boostedlow}
\end{equation}
One might have objections to this standard. Since if Eq.~(\ref{eq-lowenergy}) is satisfied, the information itself is not forming a black hole. However there is always some $\gamma$ large enough to invalidate Eq.~(\ref{eq-boostedlow}). So what is the physical meaning of this frame-dependent standard?

The answer is similar to the logic we demonstrated in Sec.\ref{sec-fit}: such frame dependent standard is related to the practical observability. We should ask that whether a group of observers in the frame of this slice (confined to the same causal patch that forces us to choose this slice) can read the relevant information. In Fig.\ref{fig-gen}-right, that means a group of observers following a shrinking shell with a boost $\gamma$. The quanta they read in their frame will have wavelengths $(\lambda/\gamma)$, that means they need to carry $N$ units of such size to keep those information. Now coming back to the rest frame, it means that this group of observers will be carrying $N$ quanta of wavelength $(\lambda/\gamma^2)$. They will form a black hole unless 
\begin{equation}
N \frac{\gamma^2}{\lambda} \ll L~,
\end{equation}
which is exactly the same standard as Eq.~(\ref{eq-boostedlow}). Therefore, Eq.~(\ref{eq-lowenergy}), applying to quantities in the frame of a slice, is the appropriate standard to determine whether information on such slice is legal within low-energy theories.

Now we have established the standard, it is straightforward to show that neither thermalization nor confinement changes our conclusion in Sec.\ref{sec-Sch}. Thermalization changes $N$ and $\lambda$ together proportionally, so Eq.~(\ref{eq-lowenergy}) is not affected at all. Confining Hawking radiation into a region of size $L$ around the black hole can reduce the boost factor in Eq.~(\ref{eq-boost}), but such change is exactly canceled by the explicit presence of $L$ in Eq.~(\ref{eq-boostedlow}).
\begin{equation}
N = M^2 \ll \frac{\lambda}{\gamma}\frac{L}{\gamma}
=\frac{M}{\sqrt{L/\Delta t}}\frac{L}{\sqrt{L/\Delta t}}
=M\Delta t~.
\end{equation}
This is exactly the same as Eq.~(\ref{eq-lR}), so our conclusion remains unchanged.

\section{Discussion}
\label{sec-dis}

\subsection{Information paradox and the quantum second law}

We have verified that a distillation-free version of the AMPS paradox does not exist. It also seems to be clear that if a distillation process is possible, then the paradox does exist. Since such a process allows us to extract the relevant information into a few qubits $R_B$ instead of keeping track of the entire $R$, this much smaller number of quanta is obviously quite mobile and can likely be further operated to satisfy the standard of weak complementarity.

On the other hand, the standard picture of distillation processes has it own problems. For example, it might take a very long time \cite{HH} or result in back-reactions \cite{HuiYan13}, and both concerns can neutralize the paradox. In fact, our discovery here works coherently with the back-reaction argument. We showed that without a distillation process, $R$ is strictly in the past of $A$. Since a distillation process acts on $R$, by causality it can of course affect $A$.

Still, we understand that the arguments involving a distillation process cannot be easily resolved unanimously. The true problem is that such a process is intrinsically unknown. It depends on how the information is encoded in $R$, which depends on the unknown black hole S-matrix. Our work can also provide a possibility to improve such situation. We can connect the information paradox to another type of distillation process which is not related to the unknown $S$-matrix. 

In order to make such connection, first note that our generalizations in Sec.\ref{sec-gen} involves thermalization: take $N$ Hawking quanta of wavelength $\lambda$ and split them into $\alpha N$ quanta with wavelength $\alpha\lambda$. We assume that the information has to become hidden in all $\alpha N$ quanta, which is essential to maintain our argument. The paradox can be restored if we discard such assumption. 

In other words, one can look for a distillation process that increases the information/energy ratio, namely some form of information density. If there is a general process in quantum thermodynamics that takes energy away without taking information away, then we can increase such information density. This will allow the information carrier to have arbitrarily low energy and avoid the problem we pointed out. Alternatively, one can interpret our work as claiming that if there is no information paradox, then there must be an in-principle obstruction against reducing quantum information density. At least, we should not be able to reduce the energy of a qubit of information from $M$ to an arbitrarily low value within time $M^3$.

\subsection{Causal patch complementarity}

The two information flows in Fig.\ref{fig-flow} base on the same principle. However, only the situation for an outside observer was widely appreciated before, and it might be misleading in some way. It is easy to believe that the black hole horizon has the mysterious UV property to guide the information. The AMPS paradox and our work together serves as a reminder that from the very beginning, the black hole horizon is not special. It only looks special for outside observers, because it coincides with the boundary of their causal patches. If we keep in mind that all boundaries of causal patches can be similarly special, then the paradoxes can be avoided.

This realization may have a profound consequence. For outside observers, going through the black hole horizon is dropping to the ``inside'' of something, and it is natural to believe that information eventually comes back. However, in most cases the causal patch boundary leads to ``outside'', and there are rarely good reasons for the information to come back from there. The late, in-falling patch is a good example showing that if we look hard enough, then we might discover a reason that information must come back instead of flowing ``outside''.

This makes us wonder whether we should also look harder in other situations, like for cosmological horizons. We just eliminated the na\"ive distinction between information going ``inside'' and ``outside'', so it becomes less crazy to think that some information leaving a cosmological horizon is actually not lost. Up to this moment, all these unexpected information flows have been demanded by the necessity to avoid paradoxes. It will be fascinating if one can come up with a more direct condition to determine whether a causal patch boundary retains information or not.

\acknowledgments

We thank Raphael Bousso, Ben Freivogel, Daniel Harlow and Erik Verlinde for useful discussions. This work is supported in part by the Foundation for Fundamental Research on Matter (FOM), which is part of the Netherlands Organization for Scientific Research (NWO).

\bibliographystyle{utcaps}
\bibliography{all}

\end{document}